\documentclass[conference]{IEEEtran}
\IEEEoverridecommandlockouts
\usepackage{cite}
\usepackage{amsmath,amssymb,amsfonts}
\usepackage{algorithmic}
\usepackage{graphicx}
\usepackage{textcomp}
\usepackage[table]{xcolor}
\usepackage{threeparttable}

\usepackage{comment}
\usepackage{kotex}
\usepackage{multirow}
\usepackage[colorlinks]{hyperref}
\hypersetup{colorlinks=true, urlcolor=blue, linkcolor=blue, citecolor=blue}
\usepackage[capitalise, nameinlink]{cleveref}
\crefformat{section}{#2\S#1#3}
\crefname{equation}{}{}
\crefname{figure}{Fig.}{Figs.}
\renewcommand\thesubsectiondis{\thesection.\arabic{subsection}}
\renewcommand\thesubsection{\mbox{\thesection.\arabic{subsection}}}
\renewcommand\thesubsubsection{\mbox{\thesection.\arabic{subsection}.\arabic{subsubsection}}}

\crefformat{appendix}{#2Appendix#3}
\crefname{equation}{}{}

\usepackage{pifont}
\newcommand{\Step}[1]{\ding{\number\numexpr181 + #1\relax\xspace}}
\newcommand{\True}{{\color{teal}\ding{52}}}
\newcommand{\False}{{\color{purple}\ding{55}}}

\usepackage{tikz}
\usetikzlibrary{shapes.geometric}
\definecolor{GraphPink}{RGB}{237,118,195}
\newcommand{\Star}[1][GraphPink]{%
    \tikz[baseline=-0.65ex]{%
        \node[
            star, 
            star points=5, 
            star point ratio=2.5,
            fill=#1,
            draw=black,
            thin,
            inner sep=0pt, 
            minimum size=.8em
        ] {};%
    }%
}

\usepackage{xspace}
\newcommand{\BfPara}[1]{{\noindent\bf#1.}\xspace}

\usepackage{booktabs}
\usepackage{subfigure}

\newcommand{\ModelName}{\textsf{DRIFT}}

\usepackage{collcell}
\usepackage{pgf}
\colorlet{MinColor}{white}
\colorlet{MaxColor}{gray!50} 
\newcommand{\ApplyGradient}[1]{
    \pgfmathparse{#1 < 0.03 ? 0 : (#1 > 0.15 ? 100 : (#1 - 0.03) * 1000)}%
    \pgfmathsetmacro{\Percent}{\pgfmathresult}%
    
    \edef\x{\noexpand\cellcolor{MaxColor!\Percent!MinColor}}\x
    #1
}
\newcolumntype{R}{>{\collectcell\ApplyGradient}r<{\endcollectcell}}

\def\BibTeX{{\rm B\kern-.05em{\sc i\kern-.025em b}\kern-.08em
    T\kern-.1667em\lower.7ex\hbox{E}\kern-.125meX}}

\newcommand{\legendsquare}[1]{\textcolor[RGB]{#1}{\rule{3ex}{1.5ex}}}

\newcommand{\Requirement}[1]{%
  \mbox{\colorbox{gray!15}{\strut \textsf{R#1}}}%
}
\begin{document}

\title{DRIFT: Drift-Resilient Invariant-Feature Transformer for DGA Detection\\
\thanks{\IEEEauthorrefmark{1}These authors contributed equally to this work.}
\thanks{We gratefully acknowledge the Cyber Analysis \& Defense department of Fraunhofer FKIE for granting us access to DGArchive~\cite{Plohmann2016DGArchive}.}
\thanks{This is the author's accepted manuscript. The version of record is published as: C.~Lee, C.~Jung, and S.~Jeong, ``DRIFT: Drift-Resilient Invariant-Feature Transformer for DGA Detection,'' in \textit{Proc. 2026 56th Annual IEEE/IFIP International Conference on Dependable Systems and Networks (DSN)}, Charlotte, NC, USA, 2026, pp.~786--799, doi: \href{https://doi.org/10.1109/DSN69566.2026.00077}{10.1109/DSN69566.2026.00077}.}
\thanks{\textcopyright~2026 IEEE. Personal use of this material is permitted. Permission from IEEE must be obtained for all other uses, in any current or future media, including reprinting/republishing this material for advertising or promotional purposes, creating new collective works, for resale or redistribution to servers or lists, or reuse of any copyrighted component of this work in other works.}
}
\author{\IEEEauthorblockN{Chaeyoung Lee\IEEEauthorrefmark{1}, Chaeri Jung\IEEEauthorrefmark{1}, and Seonghoon Jeong}
\IEEEauthorblockA{\textit{Division of Artificial Intelligence Engineering} \\
\textit{Sookmyung Women's University}\\
Seoul, Republic of Korea\\
\{amy8985, chaerry502, seonghoon\}@sookmyung.ac.kr}
}
\maketitle

\begin{abstract}
Domain Generation Algorithms (DGAs) evolve continuously to evade botnet detection, posing a persistent challenge for dependable network defense. While deep learning-based detectors achieve strong performance under static conditions, they suffer severe degradation when facing temporal drift. Through a 9-year longitudinal study (2017--2025), we empirically show that state-of-the-art character- and word-based DGA classifiers rapidly lose effectiveness as new DGA variants emerge.

To address this problem, we propose a drift-resilient Transformer-based framework that learns invariant representations through a hybrid tokenization strategy and multi-task self-supervised pre-training. The model integrates (i) character-level encoding to capture stochastic morphological patterns and (ii) subword-level encoding for word-based DGAs. Three pre-training tasks enable the model to learn robust structural and contextual features prior to supervised fine-tuning. Comprehensive evaluations demonstrate that our method significantly mitigates temporal degradation and consistently outperforms state-of-the-art baselines in forward-chaining experiments. The proposed approach offers a dependable foundation for long-term DGA defense in evolving threat landscapes. Our code is available at: https://github.com/snsec-net/2026-DSN-DRIFT.
\end{abstract}


\section{Introduction}
Domain Generation Algorithms (DGAs) are a core mechanism used by modern botnets to maintain resilient Command and Control (C\&C) communications~\cite{Manos2012ThrowAway}. By continuously generating large numbers of pseudo-random domain names, malware can evade static blacklists and preserve connectivity even when individual rendezvous points are taken down~\cite{Plohmann2016DGArchive}. Consequently, the timely and accurate identification of algorithmically generated domains remains an essential component of network defense~\cite{Mahboubi2025Evolving}. While early solutions relied heavily on manually crafted lexical or statistical features~\cite{yadav2010detecting}, deep learning (DL)-based approaches have become the \textit{de facto} standard~\cite{Saeed2022Saeed, Alqahtani2024Advances}, typically operating on character-level or word-level representations of domain strings.

Despite the short length of domain names, the DGA detection landscape is highly dynamic~\cite{Plohmann2016DGArchive}. The assumption that short strings limit variability has been disproven by recent studies~\cite{Aravena2023Dom2Vec}. Instead, the ecosystem is shaped by two opposing yet simultaneous forces: (i) attackers continually modifying generation logic to evade filters~\cite{Cebere2024DownToEarth}, and (ii) legitimate domains exhibiting increasing diversity due to new services, naming conventions, and branding trends (see~\cref{subsec:domain_name_dataset}). Together, these trends introduce substantial concept drift, undermining the static distribution assumptions underpinning most DGA datasets and detection models.

As a result, although state-of-the-art DL models report high accuracy under static evaluations~\cite{Ding2023HMT, Chen2025HDDN}, their robustness degrades sharply over time. Through a 9-year longitudinal evaluation (2017--2025), we show that the performance of existing models deteriorates rapidly as the temporal gap between training and deployment widens (see~\cref{subsec:temporal_performance_degradation}). Specifically, False Negative Rates (FNRs) surge as previously unseen DGAs emerge, while False Positive Rates (FPRs) for benign domains also increase due to evolving legitimate naming practices. This dual degradation indicates that current supervised models overfit to transient lexical features rather than learning the invariant structural cues required for long-term resilience.

To address these limitations, we propose \ModelName{}, a drift-resilient DGA detection framework that explicitly targets temporal robustness. Our method employs a dual-branch Transformer architecture that processes domain names through a hybrid tokenization strategy: one branch encodes character-level information to capture stochastic morphological patterns, while the other encodes subword-level information to model word-based DGAs. Complementing this architecture, we introduce a multi-task self-supervised pre-training phase based on three auxiliary objectives---Masked Token Prediction (MTP), Token Position Prediction (TPP), and Token Order Verification (TOV) (see~\cref{subsubsec:subtasks}). This pre-training step enables the model to learn structural regularities and contextual semantics before supervised fine-tuning, producing representations that are robust to distribution shifts.

The contributions of this paper are summarized as follows: 

\begin{itemize} 
\item We conduct a 9-year longitudinal study (2017--2025) demonstrating the severity of temporal performance degradation in state-of-the-art character-based and word-based DGA detectors. Our dataset is available at \cite{za2s-9e09}.
\item We introduce a drift-resilient detection framework that combines hybrid tokenization with multi-task self-supervised pre-training to learn invariant structural features of domain names.
\item Through extensive evaluations, we show that our framework significantly outperforms baseline models particularly in FNR over time, offering a dependable solution for long-term DGA defense.
\end{itemize}

The remainder of this paper is organized as follows. \cref{sec:preliminaries} describes the longitudinal dataset and reviews related work, motivating the need for robust DGA detection. \cref{sec:proposed_method} presents the proposed hybrid tokenization and dual-branch Transformer architecture. \cref{sec:experimental_results} provides a detailed experimental analysis and comparison against state-of-the-art baselines. Finally, \cref{sec:conclusion} concludes the paper and outlines future research directions.

\section{Preliminaries and Problem Statement}
\label{sec:preliminaries}
This section provides the foundational background for our study. We first describe the construction of the large-scale longitudinal dataset covering a 9-year period (2017--2025). Using this dataset, we then examine the temporal robustness of existing DGA detectors, highlighting the severity of concept drift and motivating the need for a drift-resilient detection framework.

\subsection{Domain Name Dataset}
\label{subsec:domain_name_dataset}
A key requirement for evaluating temporal robustness is the availability of a longitudinal dataset in which benign and malicious domains are temporally aligned. To this end, we curate domain names from multiple historical sources and aggregate them into a 9-year dataset.

\subsubsection{Benign Domains}
We source benign domains from two widely used popularity rankings: (1) the Alexa Top 1M list and (2) the Tranco Top 1M list. Historical Alexa snapshots were retrieved using the Internet Archive's Wayback Machine~\cite{website-wayback-alexa}, while historical Tranco lists were obtained through the Tranco API~\cite{Pochat2019Tranco, website-tranco}. This allows us to reconstruct year-specific benign domain distributions from 2017 to 2025.

\BfPara{Alexa Top 1M}
Alexa historically served as a \textit{de facto} benchmark for web popularity based on browser-extension telemetry. Although discontinued in 2022, its archived lists remain essential for reconstructing historical benign domain distributions.

\BfPara{Tranco Top 1M}
To mitigate manipulation and sampling biases inherent in single-source rankings, we incorporate Tranco, which aggregates rankings from multiple providers (\textit{e.g.,} Alexa, Majestic, Umbrella) using a manipulation-resistant methodology~\cite{Pochat2019Tranco}. This offers a more stable and reliable view of real-world benign domain trends.

Although both lists are intended to represent legitimate domains, prior work has shown that Top 1M datasets may contain a small proportion of malicious or parked domains~\cite{Pochat2019Tranco, Cebere2024DownToEarth}. Given the scale of our dataset and the practical challenges of large-scale sanitization, we follow standard practice and treat these lists as benign.

\subsubsection{Malicious Domains}
Malicious domains are collected from DGArchive, a comprehensive repository of DGA-generated domain names curated by the Fraunhofer Institute. This platform is built upon the foundational reverse-engineering methodology established in the seminal work by Plohmann et al.~\cite{Plohmann2016DGArchive}, which has largely defined the academic standards for DGA analysis. Rather than merely aggregating suspicious domains, DGArchive provides deterministic outputs derived directly from malware algorithms and seeds. Due to its high data transparency, a vast body of research utilizes its DGA studies and datasets as the ground truth~\cite{Drichel2023False, Cebere2024DownToEarth}. While many other DGA sources only provide domain names, we chose DGArchive because it offers precise temporal metadata required for longitudinal experiments on concept drift. Specifically, unlike passive traffic logs, DGArchive enumerates the full domain space for each malware family with per-domain timestamps, from which we extract all DGA domains active from 2017 to 2025, ensuring temporal alignment with our benign dataset. Any DGA domain names that also appeared in the Alexa or Tranco 1M lists were removed from the benign set to prevent cross-contamination.

\subsubsection{Dataset Characteristics}
\cref{table:dataset_stats} reports cumulative statistics for benign and DGA domains across 2017--2025. Although each Top 1M list contains one million entries per snapshot, the cumulative number of unique benign domains reaches approximately 49.4 million due to high churn in popular domains. This highlights a key challenge: a single-year snapshot is insufficient to represent benign naming patterns. Over a one-year period, approximately 33.13\% of the entries are replaced by new ones (\textit{cf.}\ Fig.\ 6 in~\cite{Pochat2019Tranco}). This is because domain names are inherently tied to real-world trends and shifting naming landscapes (\textit{e.g.,} the ChatGPT surge), and relying on such a snapshot risks introducing sampling bias and overfitting. This high volatility suggests that a static model trained on a single snapshot of benign data may fail to generalize as the underlying distribution of popular domains evolves over time.

The DGA dataset exhibits even larger scale, accumulating over 149 million unique malicious domains across 2017--2025. This diversity provides a comprehensive ground truth, enabling models to learn the wide spectrum of algorithmic generation strategies and their evolution over time. 
We collected 148 DGA families from the 151 provided by DGArchive, selecting those best suited for longitudinal analysis. Based on previous work~\cite{Plohmann2016DGArchive, LaO2024LLMs, Gregorio2025} and the metadata from DGArchive, we classified them into character-based and word-based types, yielding 133 and 15 families, respectively. Despite the relatively low frequency of word-based DGAs in real-world traffic, this dataset contains a significant number of word-based DGA families. Moreover, the continuous emergence of new DGA families designed for detection evasion makes shifts in data distribution and concept drift inherently inevitable.

Together, the longitudinal benign and malicious datasets allow for controlled evaluation of temporal robustness and enable the forward-chaining experiments described in \cref{subsec:performance_comparison}.

\begin{figure*}
    \centering
    \includegraphics[width=.9\linewidth]{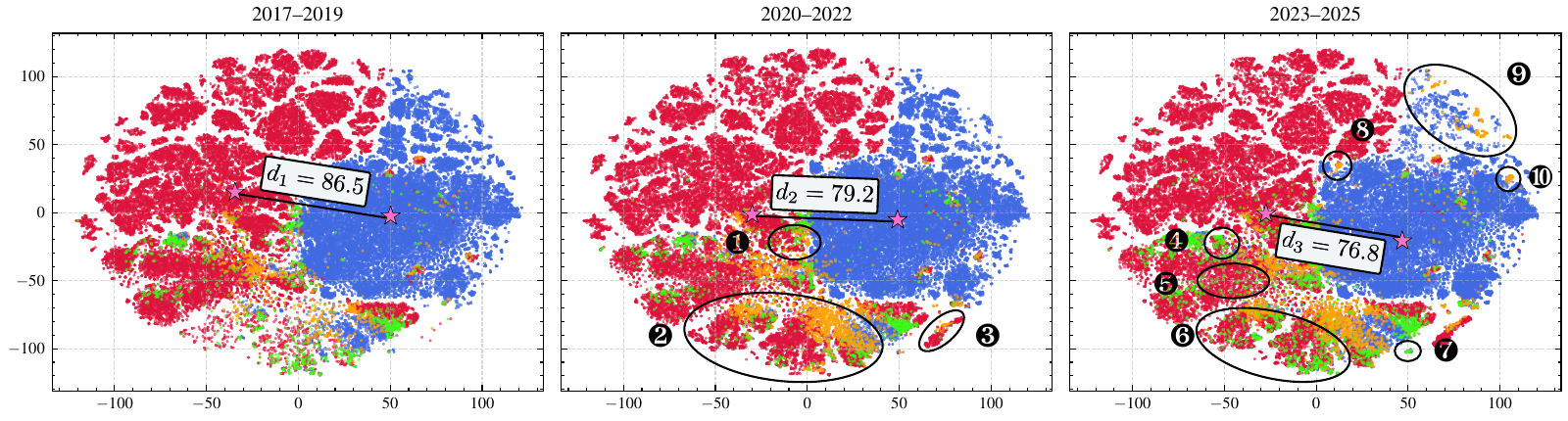}
    \vspace{-1em}
    \caption{Longitudinal t-SNE analysis of the $1024$-dimensional fused latent vector $v_{\text{fusion}}$, applied jointly to $25{,}000$ randomly sampled domains. Each point is a domain embedding, color-coded by prediction outcome---\legendsquare{220, 20, 60} TP, \legendsquare{65, 105, 225} TN, \legendsquare{57, 255, 20} FP, and \legendsquare{255, 165, 0} FN. The \Star\ symbol marks the per-class centroid (benign, DGA), and $d$ denotes the Euclidean distance between the benign and DGA centroids. To emphasize significant temporal shifts while suppressing negligible ones, we partition the 9-year dataset into three 3-year periods: 2017--2019 (train set), 2020--2022, and 2023--2025.}
    \label{fig:tsne}
    \vspace{-1em}
\end{figure*}

\subsubsection{Longitudinal Dual-sided Concept Drift}
\cref{fig:tsne} illustrates the dual-sided concept drift---both benign and DGA distributions shift over time and gradually converge to each other in the latent space---observed across our 9-year dataset. The classification outcomes shown in the figure are produced by our proposed model (detailed in \cref{sec:proposed_method}).

\BfPara{Concept Drift Over Time} The decreasing centroid distance---$d_1=86.5 > d_2=79.2 > d_3=76.8$---indicates that the benign and DGA latent spaces are converging, with each period extending further into the other's territory and raising both False Negative (FN) and False Positive (FP) cases. \Step{1}--\Step{3} and \Step{9} exhibit a substantial increase and wider spread of DGA data points. Specifically, samples in \Step{1} (\textit{e.g.,} \texttt{vutotoid}) emerged, filling the gap between the DGA and benign clusters, visibly narrowing the margin, while \Step{9} (\textit{e.g.,} \texttt{heehwaes.ddns}) shows DGA points scattered across the upper territory of the benign cluster. Both \Step{1} and \Step{9} often consist of pronounceable n-grams, appearing designed to mimic human-readable benign domains and evade detection.

\BfPara{Benign Misclassified as DGA} \Step{1} also presents interesting FP cases: a domain such as \texttt{hexieshaanxi}---a legitimate Chinese domain---is misclassified due to its unfamiliar n-gram distribution to English-centric models. Non-English domains romanized from languages whose character patterns diverge from English are inherently susceptible to such FP cases~\cite{lee2024chinese}, and longer domain lengths further push these features toward DGA-like characteristics. Acronyms in \Step{4}--\Step{6} also trigger FP cases---\textit{e.g.,} \texttt{jgpnis}, which stands for \textit{Javni Gradski Prevoz} in Ni\v{s}, Serbia. Such domains exhibit low pronounceability and sometimes contain numbers; the model misinterprets these domains due to their structural similarity to certain DGA families. Finally, in \Step{7}, FP cases are triggered by extremely short domain names consisting of only three characters---\textit{e.g.,} \texttt{mbm}, \texttt{ord}, and \texttt{hat}---which lack sufficient features for reliable classification.

\BfPara{DGA Mimicking Benign} Regarding FN cases, \Step{8} and \Step{10} demonstrate structural characteristics of evolving DGA families that can lead to misclassification. \Step{8} lies between DGA samples of mixed long alphanumeric strings and benign samples of long, readable word sequences; it contains partially pronounceable mixed alphanumeric examples (\textit{e.g.,} \texttt{jozzdd2jhftgiht}). \Step{10} further highlights DGAs (\textit{e.g.,} \texttt{osizi-deguq}, \texttt{elanat-amys}) mostly from the \texttt{Gazavat} family, whose structure appears to mimic nearby benign domains (\textit{e.g.,} \texttt{ginza-nagano}).

\begin{table}[t]
\caption{Statistics of the Collected Dataset (Cumulative from 2017)}
\vspace{-0.3em}
\label{table:dataset_stats}
\centering
\setlength{\tabcolsep}{2pt}
\resizebox{\columnwidth}{!}{%
\begin{tabular}{@{}lrrrrr@{}}
\toprule
\multirow{2}{*}{\textbf{Period}} & \multicolumn{2}{c}{\textbf{Benign domains}} & \multicolumn{2}{c}{\textbf{DGA domains}} & \multirow{2}{*}{\textbf{\begin{tabular}[c]{@{}l@{}}Uniq. DGA\\ families\end{tabular}}} \\ \cmidrule(l){2-3} \cmidrule(l){4-5}
 & \multicolumn{1}{c}{Total domains} & \multicolumn{1}{c}{Uniq. domains} & \multicolumn{1}{c}{Total domains} & \multicolumn{1}{c}{Uniq. domains} \\ \midrule
2017--2017 & 1,913,418 & 1,913,418 & 14,888,780 & 14,888,780 & 58 \\
2017--2018 & 18,729,906 & 17,129,997 & 30,004,459 & 29,305,992 & 62 \\
2017--2019 & 40,508,820 & 29,359,365 & 46,423,032 & 44,940,295 & 65 \\
2017--2020 & 58,149,929 & 36,796,092 & 65,116,780 & 62,272,656 & 71 \\
2017--2021 & 72,679,722 & 41,856,865 & 84,661,350 & 79,752,910 & 77 \\
2017--2022 & 88,262,644 & 47,358,571 & 104,424,484 & 97,107,537 & 80 \\
2017--2023 & 92,962,312 & 48,190,058 & 124,186,047 & 114,189,860 & 81\\
2017--2024 & 94,833,375 & 48,805,692 & 144,981,221 & 132,177,150 & 148 \\
2017--2025 & 96,849,575 & 49,433,110 & 165,824,441 & 149,405,584 & 148\\ \bottomrule
\end{tabular}%
}
\vspace{-2em}
\end{table}

\subsection{Temporal Performance Degradation}
\label{subsec:temporal_performance_degradation}
\begin{figure*}[!t]
    \centering
    \includegraphics[width=1\linewidth]{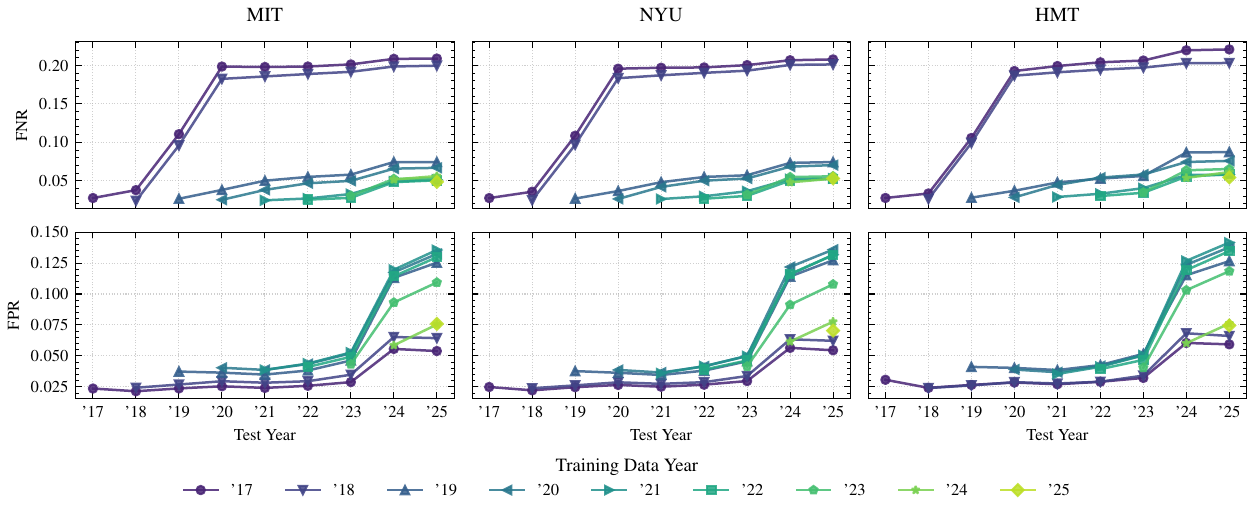}
    \vspace{-2.5em}
    \caption{Longitudinal evaluation of DGA detection performance across three baseline models (MIT~\cite{Yu2018NYU}, NYU~\cite{Yu2018NYU}, and HMT~\cite{Ding2023HMT}) over a 9-year period. The top row shows the FNR, and the bottom row shows the FPR. Each line corresponds to a model trained on a specific year and evaluated on all subsequent years, illustrating the temporal decay in detection performance caused by concept drift.}
    \label{fig:temporal_degradation}
    \vspace{-1.7em}
\end{figure*}

To quantify the impact of concept drift on existing DGA detectors, we perform a longitudinal analysis using three representative supervised models: MIT~\cite{Yu2018NYU}, NYU~\cite{Yu2018NYU}, and the Hybrid Modified Transformer (HMT)~\cite{Ding2023HMT}. We follow a forward-chaining evaluation strategy where a model trained on data from year $Y_{\text{train}}$ is evaluated on test sets from subsequent years $Y_{\text{test}}$ ($Y_{\text{train}} \le Y_{\text{test}} \le 2025$). 
Moreover, to mitigate potential biases arising from variations in dataset size and class distribution, we standardized the training set for each year. Each dataset comprises 3.0 million samples, consisting of an equal split of 1.5 million benign domains and 1.5 million DGA domains. This mirrors realistic deployment scenarios in which models must operate on continuously evolving domain distributions without immediate retraining.

\cref{fig:temporal_degradation} summarizes the temporal evolution of both FNR and FPR. The results highlight several systematic vulnerabilities in current approaches:

\BfPara{Rapid Feature Obsolescence}  
All models exhibit a sharp increase in FNR as the temporal gap $\Delta t = Y_{\text{test}} - Y_{\text{train}}$ widens. For example, models trained on 2017 data fail to detect a substantial portion of DGAs introduced in later years, producing widely diverging FNR curves. This indicates that the lexical and structural cues learned during training become obsolete as DGA families evolve.

\BfPara{Limited Benefit of Retraining}
Models trained on more recent data (\textit{e.g.,} $Y_{\text{train}}=2023$) begin with relatively low error rates but degrade rapidly when evaluated only one or two years into the future. This demonstrates that yearly retraining provides, at best, short-lived improvements and does not fundamentally address the instability caused by distribution shift.

These findings underscore that relying on transient lexical features is insufficient for long-term stability. The observed temporal fragility motivates the need for a detection framework capable of capturing invariant characteristics of DGA domains, ensuring robustness against the continuous evolution of generation algorithms.

\subsection{Related Works}
Research on DGA detection has evolved considerably over the past decade, progressing from traditional machine learning techniques to DL models and, more recently, Transformer-based architectures. Early systems relied on manually engineered lexical or statistical features extracted from domain names or Domain Name System (DNS) traffic, whereas modern approaches increasingly attempt to learn domain representations directly from raw strings.

\subsubsection{Machine Learning-based Approaches} 
Classical machine learning-based DGA detectors typically rely on hand-crafted features and traditional classifiers. Chen\textit{~et~al.}~\cite{Chen2018SVM} employed a Support Vector Machine trained on a small set of statistical domain features, while Sivaguru\textit{~et~al.}~\cite{Sivaguru2018RF} used a Random Forest model incorporating 26 lexical and linguistic attributes. Li\textit{~et~al.}~\cite{Li2019DTDBSCAN} combined a Decision Tree (J48) classifier with DBSCAN-based clustering for improved separation of DGA families. Although these methods demonstrated early success, they depend heavily on feature engineering and often require auxiliary DNS metadata~\cite{schiavoni2014phoenix, kwon2016psybog}, which limits scalability and generalization in real-world environments.

\subsubsection{Deep Learning-based Approaches}
The advent of deep neural networks enabled automated feature extraction from raw domain strings. Woodbridge\textit{~et~al.}~\cite{woodbridge2016Endgame} introduced the first Long Short-Term Memory (LSTM)-based DGA detector, inspiring subsequent LSTM variants~\cite{Akarsh2019LSTM, Selvi2021LSTM}. In parallel, Convolutional Neural Network (CNN)-based approaches emerged following their success in text classification~\cite{Zhang16NYU, kim2014CNNtext, kalchbrenner2014CNNtext}. Yu\textit{~et~al.}~\cite{Yu2018NYU} demonstrated that CNNs can effectively capture local character-level patterns for DGA detection. Hybrid architectures combining CNN and LSTM further improved feature expressiveness~\cite{Yu2018NYU, highnam2021real}. Attention-based models followed, such as Ren\textit{~et~al.}~\cite{ren2020dga}, who integrated CNN, bidirectional-LSTM, and attention mechanisms to emphasize salient regions of domain strings.

More recently, Transformer-based methods have gained traction due to their superior sequence modeling capabilities. Ding\textit{~et~al.}~\cite{Ding2023HMT} proposed the HMT, which augments Transformer encoders with character-level embeddings. Huang\textit{~et~al.}~\cite{Huang2022bert} leveraged BERT embeddings to enhance feature representations. These approaches capitalize on the contextual modeling strength of Transformers but still rely predominantly on character-level tokenization.

\subsubsection{Tokenization Strategies}
Most prior work treats domain names as raw character sequences, which limits effectiveness for word-based DGAs that contain meaningful substrings. Koh\textit{~et~al.}~\cite{Koh2018word-based} explored word-level tokenization, but their method is tailored to word-based DGAs and fails to generalize to purely stochastic ones. Recent studies have begun exploring hybrid- or subword-level tokenization~\cite{Ding2023HMT, Liew2023subword}, yet research in this direction remains sparse. A systematic framework that integrates both character-level and subword-level representations is still lacking, despite clear evidence that DGAs exhibit heterogeneous generation patterns.

\subsection{Challenges and Motivation}
In this subsection, we summarize the key challenges in drift-resilient DGA detection and derive the corresponding requirements (\Requirement1--\Requirement4) that motivate our design. Each challenge--requirement pair directly informs the architecture and evaluation strategy presented in the following sections.

\BfPara{Challenge 1: Need for a Temporal Evaluation Strategy}
As revealed by Cebere\textit{~et~al.}~\cite{Cebere2024DownToEarth}, most prior studies evaluate DGA detectors using randomly shuffled train-test splits drawn from the same time period. Such static evaluations mask the real-world consequences of distribution shift and fail to reveal how quickly models degrade when exposed to future domain distributions~\cite{koh2021wilds}.
\Requirement1: Our solution must adopt a temporally faithful evaluation protocol that mirrors deployment conditions and quantifies robustness against real-world temporal drift. We enforce this requirement using a forward-chaining evaluation setup described in \cref{sec:experimental_results}.

\BfPara{Challenge 2: Lack of Temporal Robustness in Existing Detectors}
As shown in our longitudinal experiments, even state-of-the-art DL models exhibit dramatic increases in missed detections with only modest temporal gaps. While routine retraining can alleviate this degradation, it poses three practical challenges: (1) promptly detecting that drift has occurred, (2) acquiring accurate ground-truth labels for newly emerging domains, and (3) preparing a stable, high-potential model as the retraining baseline. Consequently, retraining on recent data only delays degradation, offering no durable solution on its own.
\Requirement2: Our solution must learn invariant features that remain stable across years, enabling the detector to generalize to future, unseen DGAs without continuous retraining. We address this requirement with our multi-task self-supervised pre-training in \cref{subsec:pre-training}.

\BfPara{Challenge 3: Domain-Name-Only Detection}
A considerable portion of prior DGA detection pipelines (\textit{e.g.,}~\cite{Manos2012ThrowAway, Samuel2018FANCI, Iuchi2020Detection}) relies on auxiliary signals such as Non-Existent Domain (NXDOMAIN) bursts, DNS response anomalies, WHOIS metadata, or broader Open-Source Intelligence (OSINT)-based enrichment~\cite{Liu2023DialN}. However, these signals are often unreliable, incomplete, or entirely unavailable in practical deployment environments~\cite{Cebere2024DownToEarth}. As a result, such approaches can fail for (i) botnets whose DGAs remain dormant until the primary C\&C infrastructure is taken down, (ii) registered-DGA-style campaigns in which operators pre-register or ``age'' domains before activation, and (iii) DGAs with sufficiently low daily generation rates where attackers can feasibly register all algorithmically generated domains, producing no NXDOMAIN evidence at all.
\Requirement3: Our solution must therefore function robustly under a strictly domain-only setting, remaining effective even when NXDOMAIN-based cues are absent and OSINT sources are inconsistent or incomplete. We design and evaluate \ModelName{} under this domain-only assumption in \cref{sec:proposed_method}.

\BfPara{Challenge 4: Limitations of Pure Character-Level Tokenization}
Most existing models rely exclusively on character-level tokenization, which is insufficient for capturing the heterogeneous nature of DGAs---some rely on stochastic character sequences, while others generate word-based or hybrid patterns. Word-level tokenization alone is also inadequate, as it fails to generalize beyond word-based DGAs and introduces sparsity. Although recent work has begun exploring subword units~\cite{Ding2023HMT}, the field lacks a systematic framework that integrates both character-level and subword-level cues.
\Requirement4: Our solution must combine character-level and subword-level representations in a unified architecture capable of modeling both stochastic and word-based DGAs. We instantiate this hybrid dual-branch design in \cref{subsec:tokenization} and \cref{subsec:fine-tuning}.

\section{Proposed Method}\label{sec:proposed_method}

In this section, we introduce \ModelName{}, our proposed drift-resilient DGA detection framework that jointly models the semantic structure of subwords and the fine-grained lexical patterns of characters. 
An overview of the architecture is shown in \cref{fig:classifier}. 
\cref{subsec:tokenization} describes the hybrid tokenization pipeline, detailing how domain names are converted into character- and subword-level numerical representations. 
\cref{subsec:pre-training} presents the three self-supervised learning subtasks---MTP, TPP, and TOV---used to pre-train the dual-branch Transformer encoder. 
Finally, \cref{subsec:fine-tuning} outlines the fine-tuning procedure of \ModelName{} using longitudinal data from 2017 to 2019 to initialize a temporally robust classifier.

\subsection{Tokenization} \label{subsec:tokenization}
To prepare domain names for model processing, we convert each input string into a standardized numeric representation. This involves two steps: (i) preprocessing to normalize raw inputs and isolate meaningful lexical components, and (ii) tokenization to map the cleaned strings to character- and subword-level indices.

\subsubsection{Data Preprocessing}
Prior to tokenization, we apply a preprocessing pipeline that removes noise and enforces a consistent representation of domain names. This ensures that \ModelName{} learns intrinsic generation patterns rather than artifacts introduced during data collection.

\BfPara{Normalization and Character Filtering}
All domain names are lowercased to enforce case-insensitivity, treating strings such as \texttt{Google.com} and \texttt{google.com} equivalently.  
Following the IETF RFC 1035 specification~\cite{rfc1035}, valid domain labels consist of alphanumeric characters, hyphens (\texttt{-}), and dots (\texttt{.}).  
Although underscores (\texttt{\_}) appear in a small portion of our raw dataset, they do not conform to standard DNS naming rules and have become increasingly rare in modern domain registrations. We therefore exclude such cases from the final dataset.

\BfPara{Effective Second-Level Domain (SLD) Extraction}  
To isolate the generative component of each domain, we extract the \textit{effective SLD}, defined as the substring that excludes Top-Level Domains (TLDs) and country code TLD (ccTLD) suffixes. Because these suffixes are drawn from small, hard-coded lists in both benign and DGA domains, retaining them introduces high-frequency noise without contributing meaningful discriminative patterns.

We use a curated list of TLDs and country codes and apply the following concise rules: (i)~if the domain ends with a recognized ccTLD (\textit{e.g.,} \texttt{.co.kr}), remove the full ccTLD block; (ii)~otherwise, remove the substring to the right of the final dot (standard TLD); (iii)~the remaining substring becomes the effective SLD, with domains lacking dots used as-is.

Although the Mozilla Public Suffix List (PSL)~\cite{mozilla} provides comprehensive domain mappings, its complex syntax (wildcards, exception rules such as \texttt{*.kawasaki.jp}, \texttt{!city.kawasaki.jp}) makes reliable parsing at scale prone to noise. We therefore use a deterministic extraction logic centered on core domain structures, validated against \texttt{tldextract}~\cite{tldextract} with 96\% agreement on our 149M records. The 4\% discrepancy stems from platform SLDs such as \textit{blogspot}, \textit{wordpress}, \textit{ddns}, and \textit{myftp}; whereas \texttt{tldextract} keeps them in the registered domain, we categorize them as platform noise because they are heavily skewed toward specific classes and would let the model exploit class-correlated artifacts rather than DGA patterns.

\BfPara{Deduplication and Merging}
To construct a comprehensive benign dataset, we merge samples from Alexa and Tranco after preprocessing. Deduplication is performed using the effective SLDs to prevent over-representing frequently occurring domains. Only unique effective SLDs are retained in the final dataset.

\subsubsection{Tokenization}
\label{subsubsec:tokenization}
After preprocessing, each effective SLD is mapped to an integer sequence using two complementary tokenization strategies: (i) Character-level tokenization for fine-grained lexical patterns and (ii) Subword-level tokenization for semantic regularities in word-based DGAs.

\BfPara{Character-Level Tokenization}
For the character-based branch, the input alphabet consists of lowercase letters (\texttt{a}--\texttt{z}), digits (\texttt{0}--\texttt{9}), the hyphen (\texttt{-}), and the dot (\texttt{.}). We also include five special tokens---\texttt{[PAD]}, \texttt{[SEP]}, \texttt{[CLS]}, \texttt{[MASK]}, and \texttt{[UNK]}---resulting in a vocabulary of 43 unique IDs. Each domain is converted into a sequence of these indices, with \texttt{[CLS]} and \texttt{[SEP]} prepended and appended, respectively.

\BfPara{Subword-Level Tokenization}
The subword-branch uses the WordPiece algorithm, a widely adopted subword tokenization method in Transformer-based language models. This approach allows the tokenizer to capture meaningful morphemes when present (\textit{e.g.,} in word-based DGAs) while falling back to smaller subword units on random or pseudo-random strings. 

Choosing an appropriate vocabulary size $V$ is conceptually important to balance expressiveness and efficiency: a very small vocabulary degenerates toward character-level segmentation, whereas an overly large vocabulary may lead to sparse, poorly shared subwords. In practice, we evaluated $V \in \{500, 1000, 30522\}$ and observed only marginal performance differences (\textit{cf.}~\cref{table:ablation_study}), indicating that \ModelName{} is relatively insensitive to $V$ within this range; we therefore adopt $V=30522$ (including the same five special tokens) as a reasonable trade-off between sequence length and representation granularity.

\subsection{Self-Supervised Pre-Training} \label{subsec:pre-training}
To learn robust representations of domain names that capture both local patterns and global structural information, we employ a multi-task self-supervised learning scheme based on a Transformer encoder. Unlike standard language modeling, which typically focuses on next-token prediction or masked-token reconstruction only, our framework jointly optimizes three complementary objectives. Together, these objectives enable the model to learn (i) local semantic context over subwords or characters, (ii) the canonical ordering of tokens in valid domains, and (iii) the distinction between coherent and distorted sequences. The total loss is defined as the sum of the three task-specific losses:
\begin{equation}
\label{eq:l_total}
    \mathcal{L}_{\text{total}} = \mathcal{L}_{\text{MTP}} + \mathcal{L}_{\text{TPP}} + \mathcal{L}_{\text{TOV}}.
\end{equation}
By training on these objectives simultaneously, \ModelName{} acquires a richer understanding of both benign and DGA domain patterns.

\subsubsection{Embedding}
The embedding layer serves as the entry point for the model, converting discrete inputs into continuous vector representations. Let $V$ be the vocabulary size and $D$ be the dimension of the model. 
First, the input domain sequence is converted into a sequence of IDs, $X = \{x_1, x_2, ..., x_L\}$, where $L$ is the maximum sequence length. We utilize a learnable lookup table to map each $x_i$ to a dense vector $e_i \in \mathbb{R}^D$. To stabilize training, the weights of this embedding layer are initialized using Xavier normal initialization.

Because the Transformer encoder is permutation-invariant, we inject positional information via learnable positional embeddings. The final input representation $H_0$ is obtained by element-wise addition of token and positional embeddings:
\begin{equation}
    H_0 = \text{TokenEmbedding}(X) + \text{PositionalEmbedding}(X).
\end{equation}
The sequence $H_0$ is then fed into a multi-layer Transformer encoder to produce high-level contextual representations.

\subsubsection{Three Subtasks and Loss function}
\label{subsubsec:subtasks}

\begin{figure*}[t]
    \subfigure[Subword-based Transformer Backbone]{\includegraphics[width=0.48\linewidth]{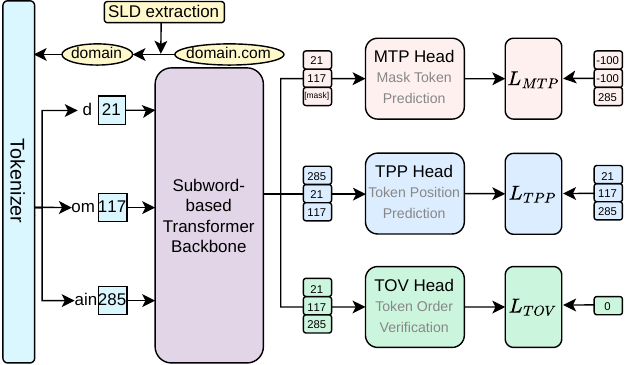}\label{subfig:subword}}
    \hfill 
    \subfigure[Character-based Transformer Backbone]{\includegraphics[width=0.48\linewidth]{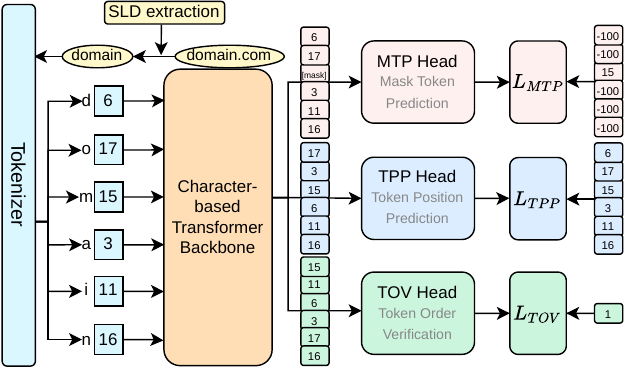}\label{subfig:char}}%
    \vspace{-0.5em}
    \caption{Proposed pre-training strategy, where each backbone processes the input domain as a sequence of character and subword tokens, respectively. To stabilize the backbone model to learn robust structural and contextual features without labeled data, we employ three auxiliary subtasks---see \cref{subsubsec:subtasks}.}
    \vspace{-1.7em}
    \label{fig:pre_training_strategy}
\end{figure*}

To capture the linguistic and structural properties of domain names, we design three auxiliary tasks depicted in~\cref{fig:pre_training_strategy}. Each task is instantiated in two variants, one for the subword-branch~\cref{subfig:subword} and one for the character-branch~\cref{subfig:char}, but they share the same underlying objective.

\BfPara{Masked Token Prediction (MTP)}  
The MTP objective comprises Masked Subword Prediction for the subword branch and Masked Character Prediction for the character-branch. This task is inspired by the Masked Language Modeling objective in BERT~\cite{Devlin2019BERT} and aims to learn bidirectional context and local dependencies. MTP is designed to teach the model which local character or subword patterns are plausible given their surrounding context.

During preprocessing, we randomly select 15\% of tokens in the input sequence and replace them with the special \texttt{[MASK]} token, yielding a corrupted sequence $\hat{X}$. The model is trained to recover the original token $x_i$ at each masked position based on the surrounding context. The loss is the cross-entropy over masked positions only:
\begin{equation}
    \mathcal{L}_{\text{MTP}} = - \sum_{i \in \mathcal{M}} \log P(x_i \mid \hat{X}),
\end{equation}
where $\mathcal{M}$ is the set of masked indices.

\BfPara{Token Position Prediction (TPP)}  
The TPP objective comprises Subword Position Prediction and Character Position Prediction. While MTP focuses on local context, TPP encourages the model to learn the global structure and correct ordering of tokens within a domain name. TPP is designed to teach the model to understand how valid domain tokens are arranged, rather than treating them as unordered bags of symbols.

In this task, we randomly shuffle the non-padding tokens of the input to obtain a scrambled sequence $X_{\text{shuffled}}$. The model receives $X_{\text{shuffled}}$ as input and must reconstruct the original sequence by predicting the correct token for each position. Unlike MTP, which computes loss only on masked tokens, TPP applies a cross-entropy loss over all valid positions:
\begin{equation}
    \mathcal{L}_{\text{TPP}} = - \sum_{t=1}^{L} \log P\big(x_t^{\text{original}} \mid X_{\text{shuffled}}\big).
\end{equation}
This objective acts as a denoising autoencoding task and promotes learning of canonical domain structures.

\BfPara{Token Order Verification (TOV)}  
The TOV objective comprises Subword Order Verification and Character Order Verification. This is a sequence-level binary classification task that trains the model to discriminate between coherent and scrambled domain names. TOV is designed to teach the model to recognize whether an entire domain looks like a realistic sequence or a structurally corrupted one.

For each batch, 50\% of sequences are kept in their original order (label $y=0$), while the remaining 50\% are randomly shuffled (label $y=1$). To obtain a fixed-size representation for classification, we aggregate the encoder outputs using both max pooling and mean pooling over the time dimension and concatenate the resulting vectors. This pooled representation is passed through a classification head to produce a probability $p$ that the input has been scrambled. The loss is the binary cross-entropy:
\begin{equation}
    \mathcal{L}_{\text{TOV}} = - \big[ y \log p + (1-y) \log (1-p) \big],
\end{equation}
where $y \in \{0, 1\}$ indicates whether the sequence order was preserved. This task encourages the model to capture high-level semantic coherence beyond token-level statistics.

\subsection{Fine-tuning for DGA Detection} \label{subsec:fine-tuning}
\begin{figure}
    \centering
    \includegraphics[width=\linewidth]{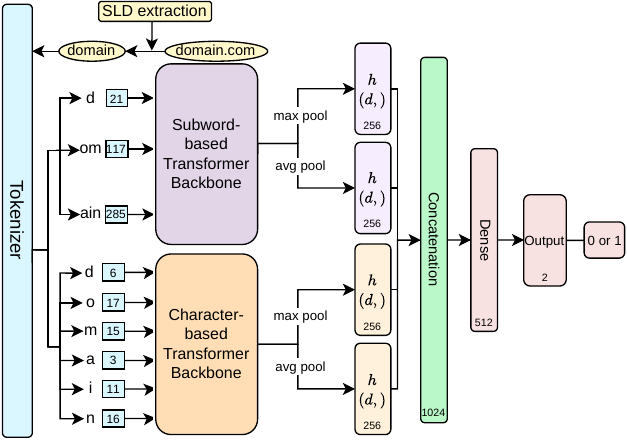}
    \vspace{-1.5em}
    \caption{Proposed dual-branch architecture for DGA detection. Subword-level and character-level Transformer encoders are pre-trained independently and then fused via pooled representations for binary classification.}
    \label{fig:classifier}
    \vspace{-1.7em}
\end{figure}

After pre-training the subword-based and character-based backbones independently, we integrate them into a unified architecture for the downstream task of DGA detection. This dual-branch design allows the model to exploit both the semantic regularities captured by subwords and the fine-grained structural patterns captured by characters.

\subsubsection{Dual-Branch Feature Extraction}
The fine-tuning model consists of two parallel branches: a Subword encoder branch and a Character encoder branch. Each branch processes the same input domain name using the weights transferred from its respective pre-trained Transformer encoder.

Rather than relying solely on the \texttt{[CLS]} token for sequence classification, we employ a pooling strategy that aggregates information from the entire sequence. For each branch we take hidden states $H \in \mathbb{R}^{L \times D}$ from the last Transformer layer. We then apply both max pooling and mean pooling along the sequence dimension. The two pooled vectors are concatenated to form a branch-specific representation: 
\begin{equation}
v_{\text{branch}} = [\text{MaxPool}(H); \text{MeanPool}(H)],
\end{equation}
where $[\,\cdot\,;\,\cdot\,]$ denotes vector concatenation. This yields a subword-based representation $v_{\text{subword}}$ and a character-based representation $v_{\text{char}}$.

\subsubsection{Feature Fusion and Classification}
To combine information across granularities, we concatenate the representations from both branches into a single fused feature vector:
\begin{equation}
    v_{\text{fusion}} = [v_{\text{subword}}; v_{\text{char}}].
\end{equation}
This fused vector is fed into a binary classification head consisting of a fully connected layer with ReLU activation, a dropout layer for regularization, and a final linear layer that outputs logits for the two classes---benign vs. DGA. Although more expressive fusion mechanisms (\textit{e.g.,} gating or cross-attention) may further improve performance, we adopt simple concatenation in this work and leave the exploration of richer fusion strategies as future work (see also the per-family analysis in \cref{subsubsec:per_family_classification}). As the classifier operates solely on this fused representation derived from the raw domain string, it is fully consistent with \Requirement3.

\subsubsection{Training Strategy}
\label{subsubsec:training_strategy}
During supervised training, we adopt a two-stage transfer learning strategy. 
In the first stage, the parameters of both pre-trained Transformer encoders are 
frozen and only the classification head is updated. This \textit{linear probing} warm-up stabilizes optimization and adapts the head to the pre-trained feature space.

In the second stage, we unfreeze the encoders and fine-tune the entire model 
end-to-end on the DGA detection task, using a smaller learning rate for the 
backbones than for the classification head. This strategy allows us to preserve 
useful linguistic and structural features from self-supervised pre-training 
while still adapting them to the downstream detection objective.

We train the model using binary cross-entropy loss with the Adam optimizer. Let $p$ 
denote the predicted probability that a domain is DGA-generated and 
$y \in \{0,1\}$ is the ground-truth label ($y=1$ for DGA, $y=0$ for benign). 
The classification loss is
\begin{equation}
    \mathcal{L}_{\text{clf}} = - \big[ y \log p + (1-y) \log (1-p) \big].
\end{equation}

\section{Experimental Results}
\label{sec:experimental_results}
In this section, we empirically evaluate \ModelName{} on the longitudinal DGA benchmark. \cref{subsec:implementation_details} details the experimental setup; 
\cref{subsec:dga_detection_performance} reports ablations and per-family analysis that isolate the contribution of hybrid tokenization, dual-branch architecture, and self-supervised pre-training;
\cref{subsec:performance_comparison} compares \ModelName{} against state-of-the-art baselines under forward-chaining evaluation across 2017--2025;
and \cref{subsec:drift_mitigation} evaluates adaptability under periodic retraining and continuous learning drift-mitigation strategies. 

\subsection{Implementation Details}
\label{subsec:implementation_details}
All experiments are conducted on an HP OMEN 45L desktop equipped with an Intel Core Ultra 9 285K (24 cores, 5.7\,GHz), 64\,GB RAM, and an NVIDIA GeForce RTX 5090 GPU. Models are implemented in PyTorch. We use the Hugging Face tokenizers library for the WordPiece tokenizer.

\subsubsection{Dataset}
\label{subsec:implementation_details_dataset}
Unless otherwise stated, we use the full benign and DGA datasets from 2017--2019 for both self-supervised pre-training and supervised fine-tuning. For each year, 300,000 samples (150,000 benign and 150,000 DGA) are held out from the full datasets as a validation set. To evaluate temporal robustness, we test on data from later years (2020--2025), \textit{i.e.,} on domains that are strictly newer than the training period. Thus, our forward-chaining evaluation protocol is consistent with \Requirement1. Furthermore, the DGA training set contains 65 DGA families, while 83 families appear only in the test set.

\subsubsection{Backbone Model}
The subword-based backbone uses an input length of $L_{\text{sub}} = 30$ and a subword vocabulary of size $V_{\text{sub}}$ (default $V_{\text{sub}} = 30{,}522$; see the ablation in~\cref{table:ablation_study}). The character-based backbone uses $L_{\text{char}} = 77$ and a character vocabulary of size $V_{\text{char}} = 43$. Both backbones share the same Transformer encoder configuration: embedding dimension $D = 256$, $N_{\text{enc}} = 12$ encoder layers, $N_{\text{head}} = 8$ attention heads, and a feed-forward dimension of $D_{\text{ff}} = 768$.

\subsubsection{Heads and Optimization}
All three pre-training heads (MTP, TPP, and TOV) are implemented as one-layer feed-forward networks.  
For MTP and TPP, the head consists of a linear layer that projects the hidden representation to the vocabulary size $V$ (subword or character, depending on the branch). The masking ratio for MTP is 15\% of the actual token length (excluding padding), with at least one token always masked. For TPP, non-padding tokens are shuffled with probability 1.0 to form the corrupted input.  
For TOV, the input is the concatenation of max-pooled and mean-pooled encoder outputs (dimension $2D$), passed through a linear layer that returns logits for the two classes (original vs.\ shuffled). The Adam optimizer with a learning rate of $10^{-4}$ was used for the pre-training.

The final DGA detection head (used in supervised fine-tuning) is a two-layer MLP: it takes the fused representation $v_{\text{fusion}} \in \mathbb{R}^{4D}$ as input, applies a hidden linear layer with size $2D$ and ReLU activation, and then a final linear layer that outputs logits for the benign and DGA classes. We use the Adam optimizer with separate learning rates: $10^{-4}$ for the classification head and $10^{-6}$ for the backbone parameters (\textit{cf.}\ the two-stage training strategy in \cref{subsubsec:training_strategy}). The batch size $B$ is set to 128 in all experiments.

\subsubsection{Training Time and Inference Throughput}
All training uses BF16 mixed precision. Each backbone is pre-trained for 2.5M steps (about 24\,h for the subword model and 29\,h for the character model), followed by 1.4M steps of supervised fine-tuning over approximately 22\,h. At inference, \ModelName{} reaches a throughput of $\lambda = 27{,}545.6$ domains per second at batch size $B = 1{,}024$. Under a constant arrival rate $\lambda$, the maximum and average queuing latencies are $L_{\max} = (B{-}1)/\lambda = 37.14$\,ms and $L_{\text{avg}} = (B{-}1)/(2\lambda) = 18.57$\,ms, both well within the widely adopted $\leq$100\,ms DNS resolution timeout~\cite{Liang2013Measuring}, indicating negligible batching overhead for inline deployment. Crucially, this latency is self-contained: \ModelName{} consumes only the raw domain string and requires neither OSINT enrichment nor NXDOMAIN telemetry, so it remains usable against dormant, pre-registered, or low-generation-rate DGAs that yield no auxiliary evidence---a decisive deployability advantage over pipelines that depend on such signals.

\subsection{DGA Detection Performance}
\label{subsec:dga_detection_performance}

\begin{table}[t]
    \caption{Experimental setup and detection performance (Micro F1-score on 2020--2025 test sets).}
    \vspace{-2em}
    \label{table:ablation_study}
    \setlength{\tabcolsep}{6.3pt}
    \begin{center}
    \begin{tabular}{@{}c|ccrc|r@{}}
    \toprule
     & \textbf{Backbone} 
     & \begin{tabular}[c]{@{}c@{}}\textbf{Pre-training}\\\textbf{Dataset}\end{tabular} 
     & \textbf{$V_{\text{sub}}$} 
     & \textbf{Unfrozen} 
     & \begin{tabular}[c]{@{}c@{}}\textbf{F1-score}\end{tabular} \\
    \midrule
     Base & H & Benign \& DGA & 30,522 & \True & 0.9646\\
     \midrule
     (a) & H & Benign only &$^\dagger$30,522 & \False & 0.9539\\
     (b) & H & Benign only & 500 & \False & 0.9564\\
     (c) & H & Benign only & 1,000 & \False  & 0.9538\\
     (d) & H & Benign only & 30,522 & \False & 0.9533\\
    \midrule
     (e) & S & Benign \& DGA & 30,522 & \True  & 0.9511\\
     (f) & C & Benign \& DGA & --- & \True & 0.9387\\
    \midrule
    (g) & H & Benign \& DGA & 30,522 & \False & 0.9585\\
    \midrule
    (h) & H & Benign only & 30,522 & \True & 0.9584\\
    \bottomrule
    \end{tabular}
    \end{center}
    \footnotesize Branch---H: Subword--Char Hybrid, \; S: Subword only, \; C: Character only
    Vocab size---$^{\mathrm{\dagger}}$Off-the-shelf BERT-base-uncased tokenizer.
    \vspace{-1.7em}
\end{table}

\subsubsection{Base Configuration and Ablation Studies}
\label{subsubsec:configuration_ablation}
To analyze the contribution of each component in \ModelName{}, we construct several variants of the base model and report the F1-score over the 2020--2025 test sets in \cref{table:ablation_study}. The ``Base'' configuration corresponds to our proposed setup: a hybrid subword--character backbone, pre-trained on both benign and DGA domains, with the backbone encoders unfrozen during supervised fine-tuning. This configuration achieves the highest F1-score among the architectural variants in \cref{table:ablation_study} (0.9646), and the remaining rows serve as ablation studies and design justifications.

\BfPara{Tokenizer and Vocabulary Size}  
Rows (a)--(d) evaluate different tokenizer choices and vocabulary sizes for the dual-branch. Row (a) uses the off-the-shelf BERT-base-uncased tokenizer trained on general natural language, while rows (b)--(d) employ tokenizers trained on domain names with vocabulary sizes $V \in \{500, 1000, 30522\}$. The performance differences are marginal (0.9533--0.9564), and all configurations remain close to the Base setting. This indicates that \ModelName{} is relatively insensitive to the exact choice of vocabulary size in this range and that domain-name-specific tokenizers do not dramatically outperform the generic BERT tokenizer in terms of aggregate F1.

\BfPara{Effect of Dual vs. Single-Branch Backbones}  
Rows (e) and (f) compare the dual-branch to single-branch variants that use only the subword backbone or only the character backbone, keeping the pre-training dataset and training strategy aligned with the Base configuration. Both single-branch models underperform the hybrid model: the subword-only variant (row (e)) reaches 0.9511, while the character-only variant (row (f)) drops further to 0.9387. This confirms that combining subword- and character-level representations is beneficial and supports \Requirement4. In \cref{subsubsec:per_family_classification}, we further show that this hybrid design yields higher micro-averaged True Positive Rates (TPRs) across diverse DGA families.

\BfPara{Backbone Fine-tuning vs. Frozen Encoders}  
Row (g) examines whether the backbones should be fine-tuned during supervised fine-tuning. Compared to the Base model (unfrozen encoders), freezing the backbone and training only the classification head (row (g)) yields a slightly lower F1-score (0.9585 vs.\ 0.9646). This result favors end-to-end fine-tuning, indicating that adapting the pre-trained features to the DGA detection objective provides a measurable benefit.

\BfPara{Impact of DGA Samples in Pre-Training}  
Finally, row (h) evaluates the role of DGA samples in self-supervised pre-training. Relative to the Base configuration (benign and DGA pre-training), training the hybrid model on benign-only data (row (h)) results in a small but consistent drop in F1-score (0.9584 vs.\ 0.9646). This suggests that including DGA domains during pre-training slightly improves downstream detection performance, helping the model better internalize patterns that are characteristic of algorithmically generated domains rather than benign traffic alone.

\subsubsection{Ablation of Pre-training Subtasks}
\label{subsubsec:subtasks_ablation}

\begin{table}[t]
    \caption{Ablation study on the three pre-training subtasks (MTP, TPP, TOV). Performance is evaluated using micro-averaged metrics on the 2020--2025 test sets.}
    \vspace{-0.3em}
    \label{table:task_ablation}
    \centering
    \setlength{\tabcolsep}{5pt} 
    \begin{tabular}{@{}cccc|rrrr@{}}
    \toprule
    \textbf{Case} & \textbf{MTP} & \textbf{TPP} & \textbf{TOV} & \textbf{Acc} & \textbf{Prec} & \textbf{Recall} & \textbf{F1-score} \\
    \midrule
    1 & \checkmark & & & 0.9436 & 0.9835 & 0.9385 & 0.9603 \\
    2 & & \checkmark & & 0.9442 & 0.9814 & 0.9402 & 0.9603 \\
    3 & & & \checkmark & 0.9473 & 0.9802 & 0.9458 & 0.9626 \\
    \midrule
    4 & \checkmark & \checkmark & & 0.9501 & \underline{0.9844} & 0.9460 & \underline{0.9647} \\
    5 & \checkmark & & \checkmark & \underline{0.9518} & \textbf{0.9857} & \underline{0.9473} & \textbf{0.9660} \\
    6 & & \checkmark & \checkmark & 0.9493 & 0.9840 & 0.9451 & 0.9641 \\
    \midrule
    7 (Base) & \checkmark & \checkmark & \checkmark & \textbf{0.9528} & 0.9818 & \textbf{0.9481} & 0.9646 \\
    \bottomrule
    \end{tabular}
    \vspace{-1.7em}
\end{table}

\begin{table*}[t]
\caption{Year-wise FPR and FNR breakdown across seven pre-training subtask combinations (2020--2025).}
\vspace{-0.3em}
\label{table:subtasks_ablation_longitudinal}
\centering
\setlength{\tabcolsep}{4.5pt} 
\begin{tabular}{@{}ccccllllllllllll}
\toprule
\multirow{2}[2]{*}{\textbf{Case}} & \multirow{2}[2]{*}{\textbf{MTP}} & \multirow{2}[2]{*}{\textbf{TPP}} & \multirow{2}[2]{*}{\textbf{TOV}} & \multicolumn{6}{c}{\textbf{False Positive Rate (FPR)}} & \multicolumn{6}{c}{\textbf{False Negative Rate (FNR)}} \\ 
\cmidrule(lr){5-10} \cmidrule(lr){11-16}
 & & & & \multicolumn{1}{c}{2020} & \multicolumn{1}{c}{2021} & \multicolumn{1}{c}{2022} & \multicolumn{1}{c}{2023} & \multicolumn{1}{c}{2024} & \multicolumn{1}{c}{2025} & \multicolumn{1}{c}{2020} & \multicolumn{1}{c}{2021} & \multicolumn{1}{c}{2022} & \multicolumn{1}{c}{2023} & \multicolumn{1}{c}{2024} & \multicolumn{1}{c}{2025} \\ \midrule

1 & \checkmark & & & 0.0287 & 0.0270 & 0.0294 & 0.0345 & 0.0842 & 0.0932 & \textbf{0.0381} & 0.0511 & 0.0578 & 0.0597 & 0.0801 & 0.0822 \\
2 & & \checkmark & & 0.0329 & 0.0313 & 0.0339 & 0.0400 & 0.0869 & 0.0956 & 0.0458 & 0.0509 & 0.0539 & 0.0567 & 0.0751 & 0.0758 \\
3 & & & \checkmark & 0.0352 & 0.0334 & 0.0361 & 0.0433 & 0.0964 & 0.1078 & 0.0407 & 0.0458 & 0.0467 & 0.0495 & 0.0701 & 0.0718 \\ \midrule
4 & \checkmark & \checkmark & & \underline{0.0276} & \underline{0.0258} & \underline{0.0280} & \underline{0.0328} & 0.0784 & 0.0869 & 0.0418 & 0.0460 & \underline{0.0464} & 0.0484 & 0.0700 & 0.0709 \\
5 & \checkmark & & \checkmark & \textbf{0.0249} & \textbf{0.0230} & \textbf{0.0257} & \textbf{0.0302} & \textbf{0.0749} & \textbf{0.0836} & 0.0414 & \underline{0.0448} & \textbf{0.0448} & \textbf{0.0465} & \underline{0.0686} & \underline{0.0699} \\
6 & & \checkmark & \checkmark & 0.0283 & 0.0265 & 0.0292 & 0.0341 & \underline{0.0764} & \underline{0.0855} & 0.0428 & 0.0467 & 0.0469 & 0.0489 & 0.0712 & 0.0724 \\ \midrule
 7 & \checkmark & \checkmark & \checkmark & 0.0328 & 0.0311 & 0.0336 & 0.0397 & 0.0875 & 0.0965 & \underline{0.0402} & \textbf{0.0445} & \textbf{0.0448} & \underline{0.0470} & \textbf{0.0657} & \textbf{0.0668} \\ 
\bottomrule
\end{tabular}
\end{table*}

To justify the necessity of the joint objective $\mathcal{L}_{\text{total}}$ in~\cref{eq:l_total}, we analyze the performance of \ModelName{} across seven combinatorial variants of our pre-training tasks. \cref{table:task_ablation} reports micro-averaged metrics across the 2020--2025 test sets, while \cref{table:subtasks_ablation_longitudinal} provides the year-wise FPR and FNR breakdown. Among single-task variants (Cases~1--3), TOV alone yields the lowest average FNR across years, confirming that sequence-level verification captures global distributional cues that transfer well to DGA detection. Among two-task combinations (Cases~4--6), Case~5 (MTP+TOV) achieves the lowest FPR across all years, while Case~7 (Base) attains the lowest FNR. This contrast reflects a well-known precision--recall trade-off: incorporating the TPP objective makes the model more aggressive at flagging DGA domains, which reduces FNR---the more critical metric in security settings where missed malicious domains pose a greater operational risk than false alarms. Accordingly, we adopt the full joint objective $\mathcal{L}_{\text{total}}$ as the Base configuration because it yields the highest Accuracy~(0.9528) and Recall~(0.9481), thereby maximizing coverage of malicious domains at the cost of only a modest FPR increase relative to Case~5.

\subsubsection{Per-Family Classification}
\label{subsubsec:per_family_classification}

\begin{figure*}[!t]
\centering
\includegraphics[width=1\linewidth]{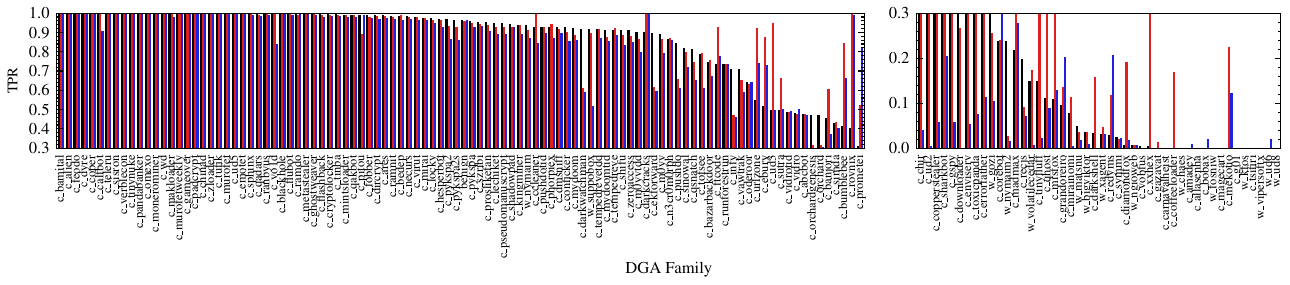}
\vspace{-2.2em}
\caption{Per-family TPRs for 147 DGA families and the benign class under three models---\legendsquare{0,0,0} Hybrid (Base), \legendsquare{255,0,0} subword-only (e), and \legendsquare{0,0,255} character-only (f). The macro-averaged TPR is 0.6713, 0.6977, and 0.6211 for Base, (e), and (f), respectively, while the micro-averaged TPR is 0.9397, 0.9201, and 0.8954. The Hybrid model outperforms these two variants on 69 out of 147 families. Overall, the Hybrid model attains the highest micro-averaged TPR, while the subword-only variant attains the highest macro-averaged TPR. For interpretability, the families are split into two panels with different y-axis ranges to make both high and low TPR regimes visible.}
\vspace{-1.4em}
\label{fig:family_tpr}
\end{figure*}

\begin{table}[t]
\caption{Representative examples of DGA families highlighting cases where the Base model achieves superior performance or, despite integrating (e) and (f), performs unexpectedly worse, supported by real domain samples.}
\vspace{-0.3em}
\label{table:case_example}
\centering
\setlength{\tabcolsep}{3pt}
\begin{tabular}{@{}l c l c c c @{}}
\toprule
\multirow{2}[2]{*}{\textbf{Family}} &
\multirow{2}[2]{*}{\textbf{Scheme}} &
\multirow{2}[2]{*}{\textbf{Examples}} &
\multicolumn{3}{c}{\textbf{TPR}} \\
\cmidrule(lr){4-6}
 &  &  & \textbf{Base} & \textbf{(e)} & \textbf{(f)} \\
\midrule
\multicolumn{6}{l}{\textit{Cases where the Hybrid model outperforms variants ($n=69/147$)}} \\
darkwatchman & c & \begin{tabular}[c]{@{}l@{}}de130b71\\ 51e88cc1\end{tabular} & 0.9190 & 0.6093 & 0.5894 \\
infy & c & \begin{tabular}[c]{@{}l@{}}7c99c1b4\\ d63bbe49\end{tabular} & 0.7094 & 0.4704 & 0.4601 \\
nymaim2 & w & \begin{tabular}[c]{@{}l@{}}runtime-incorrect\\ remedies-nude\end{tabular} & 0.2372 & 0.0228 & 0.0158 \\
\midrule
\multicolumn{6}{l}{\textit{Cases where the Hybrid model underperforms both single-branch}} \\
\multicolumn{6}{l}{\textit{variants ---(e), (f)--- ($n=15/147$)}} \\
bumblebee & c & \begin{tabular}[c]{@{}l@{}}	cmid1s1zeiu\\ tuaksrh3m4v\end{tabular} & 0.4128 & 0.8447 & 0.6616 \\
prometei & c & \begin{tabular}[c]{@{}l@{}}xinchaoagcddf\\ xinchaobhcdeg\end{tabular} & 0.3122 & 0.5258 & 0.8213 \\
urlzone & c & \begin{tabular}[c]{@{}l@{}}rwmu35avqo12tqc\\ rv1jsyatn9j3j\end{tabular} & 0.5484 & 0.9238 & 0.7437 \\
\bottomrule
\end{tabular}
\vspace{-2.1em}
\end{table}

To further examine the benefit of the dual-branch architecture, we conduct a family-level analysis over 148 classes (147 DGA families plus the benign class). Each family name is prefixed to indicate whether it is character-based or word-based. We fine-tune three model variants: (i) a subword-only variant (row (e) in \cref{table:ablation_study}), (ii) a character-only variant (row (f)), and (iii) the proposed model (Base), and compute the TPR for each class. \cref{fig:family_tpr} shows that the character-only variant is effective for character-based families but struggles with word-based DGAs, whereas the subword-only variant performs more uniformly across both types; the hybrid model achieves the highest micro-averaged TPR, benefiting from the complementary strengths of both branches.
\cref{table:case_example} illustrates both regimes: for the top three families, the subword-only and character-only variants each perform poorly but the hybrid exhibits clear synergy; for the bottom three, the single-branch variants are individually effective yet na\"ively combining them causes the hybrid to regress, suggesting that more sophisticated fusion is needed to avoid catastrophic forgetting across branches. Several of these failure cases (\textit{e.g.,}~\texttt{prometei}) involve Chinese-language contexts under-represented in training.
Overall, jointly leveraging subword and character representations is crucial for robust detection of heterogeneous DGA mechanisms, directly addressing \Requirement4.

\subsection{Forward-Chaining Performance Comparison}
\label{subsec:performance_comparison}

\begin{table}[t]
\caption{Overall forward-chaining performance on 2020--2025 test sets (micro-averaged across years).}
\vspace{-0.3em}
\label{table:performance_aprf}
\centering
\setlength{\tabcolsep}{3pt}
\begin{tabular}{@{}lrrrr@{}}
\toprule
\textbf{Method} & \textbf{Accuracy} & \textbf{Precision} & \textbf{Recall} & \textbf{F1-score} \\
\midrule
Endgame (2016)~\cite{woodbridge2016Endgame} & 0.943948 & 0.976964 & 0.939656 & 0.957947 \\
MIT (2016)~\cite{Vosoughi2016MIT, Yu2018NYU} & 0.930575 & \textbf{0.986709} & 0.910075 & 0.946844\\
NYU (2016)~\cite{Zhang16NYU, Yu2018NYU} & 0.932011 & 0.985763 & 0.913118 & 0.948051\\
B-ResNet (2020)~\cite{Drichel2020BResNet} & 0.939874 & 0.979729 & 0.930760 & 0.954617 \\
M-ResNet + B-cos (2022)~\cite{Boehle2022CVPR} & \underline{0.948813} & 0.977461 & \underline{0.946484} & \underline{0.961723} \\
Dom2Vec (2023)~\cite{Aravena2023Dom2Vec} & 0.927860 & 0.968624 & 0.923743 & 0.945651 \\
HMT (2023)~\cite{Ding2023HMT} & 0.931811 & \underline{0.986542} & 0.912077 & 0.947849\\
SFT-Llama3-8B (2024)~\cite{LaO2024LLMs} & 0.805239 & 0.895069 & 0.808071 & 0.849348 \\
HDDN (2025)~\cite{Chen2025HDDN} & 0.939153 & 0.978111 & 0.931284 & 0.954123 \\
Fine-tuned BERT~\cite{Devlin2019BERT} & 0.938252 & 0.983688 & 0.924444 & 0.953147 \\
\ModelName{} (Supervised) (\cref{fig:classifier}) & 0.941976 & 0.982171 & 0.931506 & 0.956168\\
\ModelName{} & \textbf{0.952763} & 0.981770 & \textbf{0.948077} & \textbf{0.964630} \\
\bottomrule
\end{tabular}
\vspace{-2em}
\end{table}

To evaluate the temporal robustness of DGA detection models, we conduct a comprehensive comparative analysis against ten baselines. These baselines span a wide range of methodologies, including classical DL models (Endgame~\cite{woodbridge2016Endgame}, MIT~\cite{Yu2018NYU}, NYU~\cite{Yu2018NYU}), residual networks (B-ResNet~\cite{Drichel2020BResNet}, M-ResNet~\cite{Boehle2022CVPR}), statistical embedding approaches (Dom2Vec~\cite{Aravena2023Dom2Vec}), advanced hybrid models (HMT~\cite{Ding2023HMT}, HDDN~\cite{Chen2025HDDN}), and large language models (BERT~\cite{Devlin2019BERT}, SFT-Llama3-8B~\cite{LaO2024LLMs}). All models are trained on the 2017--2019 datasets and evaluated on subsequent years (2020--2025) under the forward-chaining protocol, thereby enforcing \Requirement1 and simulating realistic model aging without retraining. 

\begin{table*}[t]
\caption{Performance comparison of existing methods across test years.}
\vspace{-0.3em}
\label{table:performance_comparison_metrics}
\centering
\setlength{\tabcolsep}{2pt}
\begin{tabular}{@{}lRRRRRRRRRRRR}
\toprule
\multirow{2}[2]{*}{\textbf{Method}} & \multicolumn{6}{c}{\textbf{False Positive Rate (FPR)}} & \multicolumn{6}{c}{\textbf{False Negative Rate (FNR)}}  \\ 
\cmidrule(lr){2-7} \cmidrule(lr){8-13}
 & \multicolumn{1}{c}{2020} 
 & \multicolumn{1}{c}{2021} 
 & \multicolumn{1}{c}{2022}
 & \multicolumn{1}{c}{2023} 
 & \multicolumn{1}{c}{2024} 
 & \multicolumn{1}{c}{2025} 
 & \multicolumn{1}{c}{2020} 
 & \multicolumn{1}{c}{2021} 
 & \multicolumn{1}{c}{2022} 
 & \multicolumn{1}{c}{2023}
 & \multicolumn{1}{c}{2024} 
 & \multicolumn{1}{c}{2025} \\ \midrule
Endgame (2016)~\cite{woodbridge2016Endgame} & 0.037587 & 0.035945 & 0.038921 & 0.046901 & 0.109103 & 0.120169 & 0.049656 & 0.059766 & 0.064794 & 0.067611 & 0.073841 & 0.075314 \\
MIT (2016)~\cite{Vosoughi2016MIT, Yu2018NYU} & 0.022765 & 0.021082 & 0.023194 & 0.027517 & 0.065128 & 0.071038 & 0.069813 & 0.081169 & 0.086732 & 0.089291 & 0.104253 & 0.105506 \\
NYU (2016)~\cite{Zhang16NYU, Yu2018NYU} & 0.024707 & 0.022929 & 0.025192 & 0.029634 & 0.066882 & 0.073718 & 0.068470 & 0.078663 & 0.083348 & 0.086131 & 0.100390 & 0.101692 \\
B-ResNet (2020)~\cite{Drichel2020BResNet} & 0.043184 & 0.041887 & 0.045256 & 0.053354 & 0.097250 & 0.108473 & 0.057026 & 0.061783 & 0.063773 & 0.066373 & 0.083641 & 0.084701 \\
M-ResNet + B-cos (2022)~\cite{Boehle2022CVPR} & 0.039640 & 0.038170 & 0.041114 & 0.048349 & 0.098267 & 0.108946 & 0.043007 & 0.048391 & 0.049730 & 0.051817 & 0.068386 & 0.068851 \\
Dom2Vec (2023)~\cite{Aravena2023Dom2Vec} & 0.050248 & 0.049740 & 0.053397 & 0.063417 & 0.127900 & 0.137200 & 0.066665 & 0.074395 & 0.078659 & 0.080442 & 0.085100 & 0.084700 \\
HMT (2023)~\cite{Ding2023HMT} & 0.023073 & 0.021395 & 0.023669 & 0.027690 & 0.065852 & 0.072185 & 0.069080 & 0.079517 & 0.085153 & 0.087280 & 0.101229 & 0.102667 \\
SFT-Llama3-8B (2024)~\cite{LaO2024LLMs} & 0.199478 & 0.200564 & 0.196810 & 0.199898 & 0.200038 & 0.200178 & 0.190985 & 0.192171 & 0.189474 & 0.192437 & 0.192921 & 0.193405 \\
HDDN (2025)~\cite{Chen2025HDDN} & 0.045432 & 0.044082 & 0.047069 & 0.056879 & 0.098528 & 0.108187 & 0.055095 & 0.064236 & 0.070306 & 0.073623 & 0.079726 & 0.080943 \\
Fine-tuned BERT~\cite{Devlin2019BERT} & 0.026280 & 0.024604 & 0.026919 & 0.031989 & 0.107327 & 0.118348 & 0.056094 & 0.068145 & 0.074343 & 0.077104 & 0.084107 & 0.091110 \\
\ModelName{} (Supervised) (\cref{fig:classifier}) & 0.031638 & 0.030062 & 0.032776 & 0.038314 & 0.081483 & 0.089660 & 0.054527 & 0.061111 & 0.063187 & 0.066023 & 0.081593 & 0.082249\\ 
\ModelName{} & 0.032849 & 0.031163 & 0.033638 & 0.039797 & 0.087579 & 0.096520 & 0.040287 & 0.044555 & 0.044820 & 0.047055 & 0.065754 & 0.066817\\
\bottomrule
\end{tabular}%
\vspace{-1em}
\end{table*}

We adopted public source code for MIT, NYU, B-ResNet, and M-ResNet~\cite{Yu2018NYU, Drichel2023False}. Given the large scale of our training set, we trained these models extensively until convergence. We used Hugging Face's Transformers~\cite{wolf2020transformers} to implement a fine-tuned BERT that works as a binary classifier---\textit{cf.} Fig.~4(b) in~\cite{Devlin2019BERT}. For SFT-Llama3-8B, we used the following prompt---``Analyze the following domain name and classify if it is generated by a Domain Generation Algorithm (DGA) or if it is a Benign domain.$\backslash$nDomain: \{domain\}$\backslash$nLabel: \{label\},'' where label $\in \{\textrm{Benign}, \textrm{DGA}\}$. We manually reproduced Endgame, Dom2Vec, HMT, and HDDN.

\begin{figure*}[!t]
\centering
\includegraphics[width=.98\linewidth]{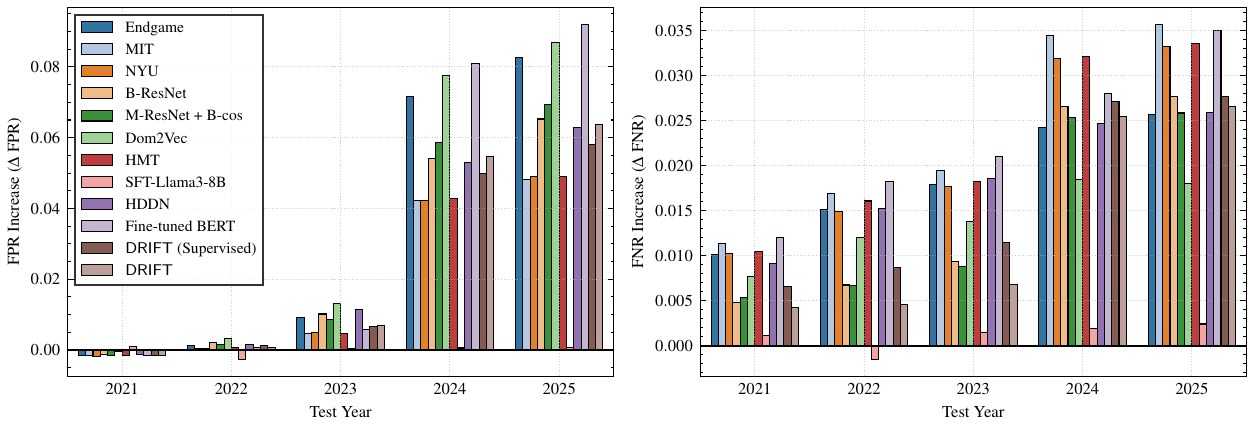}
\vspace{-1.2em}
\caption{Temporal evolution of detection errors under forward-chaining evaluation. Left: year-wise FPR for all methods. Right: year-wise FNR for all methods (2020--2025). The proposed method substantially reduces FNR drift while still exhibiting the benign FPR drift observed across all models.}
\vspace{-0.9em}
\label{fig:forward_chaining_comparison}
\end{figure*}

\cref{table:performance_aprf} reports the overall forward-chaining metrics averaged across years, while \cref{table:performance_comparison_metrics} and \cref{fig:forward_chaining_comparison} detail the year-wise FPR and FNR. Finally, \cref{table:performance_unseen} presents the micro-averaged FNR specifically for unseen DGA families across 2020--2025.

\BfPara{Vulnerability of Existing Models to Concept Drift}
As expected, most baseline models exhibit noticeable temporal degradation. Early DL models (Endgame, MIT, NYU) show a clear FNR increase as the test year moves away from training, indicating failure to generalize to later DGA families. More recent architectures (Dom2Vec, HDDN) mitigate this only partially: both FPR and FNR steadily grow over time. SFT-Llama3-8B, despite its large capacity, suffers from persistently high error rates (FPR $\approx 20\%$, FNR $\approx 19\%$), suggesting that generic generative pre-training does not directly yield the discriminative robustness required for DGA detection.

\BfPara{Impact of Benign Concept Drift on FPR}
A central observation from 2024--2025 is that \emph{all} methods experience an FPR increase---for example, Endgame and HDDN see FPR more than double, from $\sim$4\% in 2020 to $\sim$11--12\% in 2025. This benign drift reflects the evolution of legitimate domain naming practices (\textit{e.g.,} creative branding, new services); models trained on older benign distributions increasingly misclassify novel legitimate domains. Our method is not immune---FPR rises from $\sim$3\% (2020) to $\sim$9.6\% (2025)---though still below most baselines. Temporal drift on benign traffic remains an open challenge.

\BfPara{Robustness and Remaining Limitations}
Despite this residual FPR drift, \ModelName{} achieves the best overall forward-chaining performance in \cref{table:performance_aprf}, with the highest Accuracy (0.9528), Recall (0.9481), and F1-score (0.9646). More importantly from a dependability perspective, our FNR stays below 6.7\% (0.040--0.067) across all years with labeled DGAs, whereas competitive baselines such as M-ResNet+B-cos, HMT, and fine-tuned BERT exhibit FNRs in the 4--11\% range and SFT-Llama3-8B exceeds 18\%. This low and temporally stable FNR means DGA domains are reliably flagged years after deployment---critical for the dependability of the protection system.

\begin{table}[t]
\caption{Unseen-Families FNR on 2020--2025 Test Sets (Micro-Averaged)}
\vspace{-0.3em}
\label{table:performance_unseen}
\centering
\setlength{\tabcolsep}{3pt}
\begin{tabular}{@{}c|rrrr@{}}
\toprule
\textbf{Metric} & \textbf{MIT}~\cite{Vosoughi2016MIT, Yu2018NYU} & \textbf{NYU}~\cite{Zhang16NYU, Yu2018NYU} & \textbf{HMT}~\cite{Ding2023HMT} & \textbf{\ModelName{}} \\
\midrule
FNR & 0.279048 & 0.255106 & 0.259630 & 0.143913\\
\bottomrule
\end{tabular}
\vspace{-1em}
\end{table}

\BfPara{Performance Against Future Unseen DGAs} As noted in \cref{subsec:implementation_details_dataset}, 65 DGA families appear in the training set, while the remaining 83 families are exclusive to the test sets. \cref{table:performance_unseen} shows that \ModelName{} achieves the lowest FNR (0.1439) on these unseen families among the evaluated baselines (MIT, NYU, and HMT), demonstrating superior generalization to previously unseen DGA families.

In summary, although \ModelName{} is not immune to benign concept drift, it substantially reduces and stabilizes the FNR compared to existing methods, fulfilling \Requirement2 under forward-chaining evaluation and leaving further FPR reduction as future work.

\subsection{Performance under Adaptive Drift-Mitigation Strategies}
\label{subsec:drift_mitigation}
\begin{figure}
    \centering
    \includegraphics[width=\linewidth]{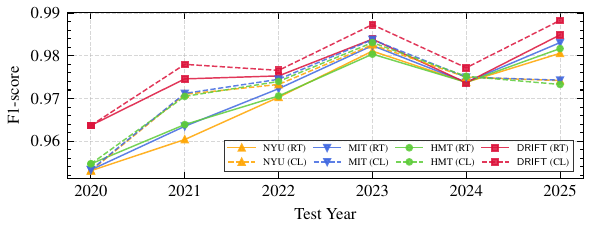}
    \vspace{-2.2em}
    \caption{Performance comparison of \ModelName{} and three baseline models under adaptive drift-mitigation strategies. Solid lines denote periodic retraining (RT) from scratch on accumulated historical data; dashed lines denote continuous learning (CL) via sequential single-epoch updates on each new year's data.}
    \vspace{-1.7em}
    \label{fig:drift_mitigation}
\end{figure}

To evaluate adaptability against continuous concept drift, we compare \ModelName{} with the three baselines (NYU, MIT, and HMT) under two adaptive drift-mitigation strategies: periodic retraining (RT) and continuous learning (CL). Both protocols are implemented at yearly granularity to align with the benchmark's year-indexed test sets and with DGArchive's annual label cadence; the same procedure applies to the remaining baselines and is omitted for brevity.

\BfPara{Periodic Retraining}
Periodic retraining trains an independent model entirely from scratch for each evaluation year, using the progressively accumulated dataset $Y_{\text{train}} \in [2017, N]$ for $2019 \le N \le 2024$, with the corresponding test year $Y_{\text{test}} = N + 1$.
For \ModelName{}, this explicitly requires repeating both the pre-training and the fine-tuning phases for every new model.

\BfPara{Continuous Learning}
Continuous learning starts from the base model trained on 2017--2019. For each $N \in \{2020, \dots, 2024\}$, the model from the previous step ($N-1$) is updated for a single epoch using only data from $Y_{\text{train}} = N$; the updated model is then evaluated on $Y_{\text{test}} = N + 1$ and serves as the base for the next iteration.
To prevent catastrophic forgetting and maximize efficiency in \ModelName{}, we freeze the pre-trained backbones and update only the classification head.

As shown in \cref{fig:drift_mitigation}, both adaptation strategies mitigate concept drift, and CL generally outperforms RT across all models. Crucially, CL-\ModelName{} (dashed red) attains the highest F1 in every test year from 2020 to 2025. This performance gain is obtained even though \ModelName{} updates only the classification head for a single epoch, whereas the baselines fully retrain their architectures in both RT and CL. This indicates that the invariant features learned through the pre-training already encode most of the cues needed to track emerging DGAs, so adaptation can be concentrated at the decision boundary. While both RT and CL incur practical costs---particularly the continuous acquisition of up-to-date labels---these results reinforce \Requirement2: when adaptation is unavoidable, \ModelName{} serves as a markedly more efficient and robust backbone than the evaluated baselines.

\section{Discussion and Conclusion}
\vspace{-0.1em}
\label{sec:conclusion}
Our 9-year longitudinal study (2017--2025) under forward-chaining evaluation shows that state-of-the-art DGA detectors degrade rapidly as the training--deployment gap widens, especially in the form of rising FNRs. These observations motivated four requirements for drift-resilient DGA detection: temporally faithful evaluation, temporal robustness, domain-name-only operation, and hybrid character/subword representations.

To address these challenges, we proposed \ModelName{}, a dual-branch Transformer that processes domain names through subword and character tokenizations and is pre-trained with three self-supervised objectives (MTP, TPP, and TOV), designed to learn invariant structural features directly from raw strings, without auxiliary signals such as NXDOMAIN bursts or OSINT enrichment. The resulting fused representation is fine-tuned for binary DGA detection under a domain-only setting, satisfying \Requirement3 and \Requirement4.

Across ten competitive baselines, \ModelName{} achieves the best overall forward-chaining performance (highest Accuracy, Recall, and F1-score averaged over 2020--2025). At the family level, the hybrid model attains the highest TPR on more families than either single-branch variant, and the highest micro-averaged TPR overall. Most importantly from a dependability standpoint, \ModelName{} substantially reduces and stabilizes the FNR over time, maintaining low FNR even for DGAs appearing years after training---indicating that our self-supervised dual-branch design generalizes to unseen DGAs, fulfilling \Requirement2.

At the same time, our results reveal that benign concept drift remains a significant challenge: like all evaluated methods, \ModelName{} experiences a non-negligible increase in FPR during 2024--2025. Reducing this benign FPR drift while preserving the low FNR remains an important direction for future work---for instance, via calibrated uncertainty, adaptive thresholding, or lightweight online adaptation that limits labeling cost. We hope our longitudinal evaluation protocol, dataset curation, and representation learning framework will serve as a foundation for future research on dependable, drift-resilient DGA detection.

\label{end:body}\typeout{BODY-END-PAGE: \pageref{end:body}}
\bibliographystyle{IEEEtran}
\bibliography{references}
\end{document}